\documentstyle[12pt]{article}
\newcommand{\boldm}[1]{\mbox{\boldmath $#1$}}
\newcommand{\mt}{m_{\tau}}

\newcommand{\spsi}{\sin\psi}
\newcommand{\cpsi}{\cos\psi}

\newcommand{\cpsiz}{\cos^{2}\psi}
\newcommand{\szpsi}{\sin 2\psi}

\newcommand{\sth}{\sin\beta}
\newcommand{\sthz}{\sin^{2}\beta}
\newcommand{\szth}{\sin 2\beta}

\newcommand{\cth}{\cos\beta}
\newcommand{\cthz}{\cos^{2}\beta}
\newcommand{\schi}{\sin\gamma}
\newcommand{\cchi}{\cos\gamma}

\newcommand{\szchi}{\sin 2\gamma}
\newcommand{\czchi}{\cos 2\gamma}

\newcommand{\ke}{\boldm{K_{1}}}
\newcommand{\kz}{\boldm{K_{2}}}
\newcommand{\kd}{\boldm{K_{3}}}
\newcommand{\kv}{\boldm{K_{4}}}
\newcommand{\kf}{\boldm{K_{5}}}

\newcommand{\keb}{\boldm{\overline{K}_{1}}}
\newcommand{\kzb}{\boldm{\overline{K}_{2}}}
\newcommand{\kdb}{\boldm{\overline{K}_{3}}}
\begin{document}
\everymath={\displaystyle}
\thispagestyle{plain}
  \def\thebibliography#1{{\bf{References}}\list
 {[\arabic{enumi}]}{\settowidth\labelwidth{[#1]}\leftmargin\labelwidth
   \advance\leftmargin\labelsep
   \usecounter{enumi}}
   \def\newblock{\hskip .11em plus .33em minus -.07em}
   \sloppy
   \sfcode`\.=1000\relax}
  \let\endthebibliography=\endlist
\everymath={\displaystyle}
\thispagestyle{empty}
\noindent
\hfill TTP92/33\\
\mbox{}
\hfill  December 1992   \\   
\vspace{0.5cm}
\begin{center}
  \begin{Large}
  \begin{bf}
MEASURING THE WESS-ZUMINO ANOMALY IN TAU DECAYS   \\
  \end{bf}
  \end{Large}
  \vspace{0.8cm}
  \begin{large}
   R. Decker and E.\ Mirkes\\[3mm]
    Institut f\"ur Theoretische Teilchenphysik\\
    Universit\"at Karlsruhe\\
    Kaiserstr. 12,    Postfach 6980\\[2mm]
    7500 Karlsruhe 1, Germany\\
  \end{large}
  \vspace{5.8cm}
  {\bf Abstract}
\end{center}
\begin{quotation}
We propose to measure the Wess-Zumino anomaly contribution by
considering angular distributions in the decays
$\tau\to\nu_\tau \eta\pi^{-}\pi^{0}$,\\
$\tau\to\nu_\tau K^- \pi^{-} K^+ $ and $\tau\to\nu_\tau K^- \pi^-
\pi^+$.
Radial excitations of the $K^*$, which cannot be seen in $e^++e^-$,
 should
be observed in the $ K^- \pi^+ \pi^-$ decay channel.
\end{quotation}
\newpage
\setcounter{page}{2}
\section{Introduction }
With the experimental progress in $\tau$-decays an ideal tool for
studying strong interaction physics has been developed. In this
paper
we show that several decays can be used to test the Wess-Zumino
anomaly \cite{WZ}. It appears that the anomaly violates the rule
that the weak axialvector- and vector-currents produce an odd and
an even
number of pseudoscalars, respectively.
{}From its structure it can be seen that the anomaly contributes
possibly to $\tau$ decays into $\nu_\tau\ +\ n$ mesons (with $n\geq 3$).
Of course the goldenplated decay is $\tau\to\nu_\tau
\eta \pi^-\pi^0$ which has a vanishing contribution from the
axialvector-current \cite{kramer,pich3,demi}. Therefore a detected
 $\eta\pi^-\pi^0$ final state implies a nonvanishing anomaly.
A recent measurement of its width \cite{CLEO} confirms the CVC
predictions
\cite{pich3,demi}. In this paper we will
present the most general angular distribution
of the $\eta\pi^{-}\pi^{0}$ system  in terms of the vector current
formfactor for this channel. Furthermore
we demonstrate that also other
  decays modes into
three pseudoscalars can be used to confirm (not only qualitatively)
the presence of the Wess-Zumino anomaly. However, the most prominent
 decay
channel into three pseudoscalars, i.e. $\pi^- \pi^- \pi^{+}$,
 cannot be used since
$G$-parity forbids the anomaly to contribute.
Beside the $\eta\pi^-\pi^0$ channel interesting candidates are the
decay
channels $K^-\pi^{-}K^+$ and
$K^-\pi^-\pi^+$. Recently the corresponding branching ratios
have been reconsidered \cite{demi} and it appears that
the branching ratios alone are not sufficient to determine the
 presence
of the anomaly.
In the course of this paper we will show that  a detailed  study of
angular
distributions, as defined in \cite{kumi1,kumi2}, is  well suited to
extract the contribution of the anomaly.\\
Our paper is organized as follows:

In section \ref{haupt} we introduce
the kinematical parameters which are adapted to  the present
experimental
situation
where the direction of flight of the tau lepton can not be
reconstructed and only the hadrons are detected.
Then we present, following \cite{kumi1,kumi2},
 the most general angular
distribution of the three hadrons in terms of hadronic structure
functions.
The dependence of the $\tau$-polarization is included.
 By considering adequate moments in section \ref{diss}
we show that all of
the hadronic structure functions can be measured without reconstructing
 the
$\tau$-restframe.
 Section \ref{form} is devoted
to the hadronic model \cite{demi}
encoded in the structure functions.
We present explicit parametrizations of the formfactors for the
 decays into
 $\eta\pi^-\pi^0$,
 $K^-\pi^{-}K^+$ and $K^-\pi^-\pi^+$. The different parameters of
the model have been moved to appendix A.

Finally numerical results for the hadronic structure functions
 of the considered channels are presented in section \ref{numer},
proving that an experimental determination of the anomaly is feasible.
We anticipate our results and
 urge experimentalists to analyse the $K^-\pi^-\pi^+$-channel
which could contain radial excitations of the $K^*$ which can not be
obtained in $e^+e^-$ experiments.\\
\section{Lepton Tensor and Angular Distributions}
\label{haupt}
Let us consider the following $\tau$-decay
  \begin{equation}
\tau(l,s)\rightarrow\nu(l^{\prime},s^{\prime})
+h_{1}(q_{1},m_{1})+h_{2}(q_{2},m_{2})+h_{3}(q_{3},m_{3})
 \label{process}
  \end{equation}
where $h_i(q_i,m_i)$ are pseudoscalar mesons.
The  matrix element reads as
  \begin{equation}
{\cal{M}}=\frac{G}{\sqrt{2}}\,
\bigl(^{\cos\theta_{c}}_{\sin\theta_{c}}\bigr)
\,M_{\mu}J^{\mu}
\label{mdef}
  \end{equation}
with $G$ the Fermi-coupling constant.
The  cosine and the sine of the Cabbibo angle ($\theta_C$)
in (\ref{mdef}) have to be used for Cabibbo allowed $\Delta S=0$ and
Cabibbo suppressed $|\Delta S|=1$ decays, respectively.
The leptonic ($M_\mu$) and hadronic ($J^\mu$) currents are given by
  \begin{equation}
M_{\mu}=
\bar{u}(l^{\prime},s^{\prime})\gamma_{\mu}(1-\gamma_{5})u(l,s)
  \end{equation}
and
  \begin{equation}
J^{\mu}(q_{1},q_{2},q_{3})=\langle h_{1}(q_{1})h_{2}(q_{2})h_{3}(q_{3})
|V^{\mu}(0)-A^{\mu}(0)|0\rangle
\label{hadmat}
  \end{equation}
$V^\mu$ and $A^\mu$ are the vector and axialvector quark currents,
respectively.
The most general ansatz for the matrix element of the
quark current $J^{\mu}$  in (\ref{hadmat})
is characterized by four formfactors \cite{kumi2,demi}.
  \begin{eqnarray}
J^{\mu}(q_{1},q_{2},q_{3})
&=&   V_{1}^{\mu}\,F_{1}
    + V_{2}^{\mu}\,F_{2}
    +\,i\, V_{3}^{\mu}\,F_{3}
    + V_{4}^{\mu}\,F_{4}\label{f1234}
  \end{eqnarray}
with
  \begin{equation}
    \begin{array}{ll}
V_{1}^{\mu}&=q_{1}^{\mu}-q_{3}^{\mu}-Q^{\mu}
\frac{Q(q_{1}-q_{3})}{Q^{2}}
\\[2mm]
V_{2}^{\mu}&=q_{2}^{\mu}-q_{3}^{\mu}-Q^{\mu}
\frac{Q(q_{2}-q_{3})}{Q^{2}}
\\[2mm]
V_{3}^{\mu}&=
\epsilon^{\mu\alpha\beta\gamma}q_{1\,\alpha}q_{2\,\beta}
                                            q_{3\,\gamma}
\\[2mm]
V_{4}^{\mu}&=q_{1}^{\mu}+q_{2}^{\mu}+q_{3}^{\mu}\,=Q^{\mu}
    \end{array}
  \end{equation}
The Wess-Zumino anomaly
 which is of main interest in the present paper give
rise to the term proportional to $F_{3}$.
The terms proportional to $F_{1}$ and $F_{2}$
 originate from the axialvector-current.
Together they correspond to a spin one hadronic final state
  while the $F_{4}$ term
is due to  the spin zero part of the axialvector-\- current.
 As it has been shown in
\cite{demi} the spin zero contributions  are extremely small
and we neglect them in
the rest of this paper, i.e. $F_{4}$ is set equal to zero.

The differential decay rate  is obtained from
  \begin{equation}
d\Gamma(\tau\rightarrow \nu_\tau\,3h)=\frac{1}{2\mt}
\frac{G^{2}}{2}\,\bigl(^{\cos^2\theta_{c}}_{\sin^2\theta_{c}}\bigr)\,
\left\{L_{\mu\nu}H^{\mu\nu}\right\}
\,d\mbox{PS}^{(4)}
 \label{decay}
  \end{equation}
where $L_{\mu\nu}=M_\mu (M_\nu)^\dagger$ and
$H^{\mu\nu}\equiv J^{\mu}(J^{\nu})^{\dagger}$.

Reaction (\ref{process})
 is most easily analyzed in the hadronic rest frame
$\vec{q}_{1}+\vec{q}_{2}+\vec{q}_{3}=0$.
The orientation of the hadronic
system is characterized by
three Euler angles ($\alpha,\beta$ and $\gamma$) introduced in
\cite{kumi1,kumi2}.
In current $e^++e^-$ ($\to \tau^+\tau^-(\to \nu_\tau 3$ mesons))
experiments two out of the three Euler angles are measurable.
The measurable ones  are defined by
 \begin{eqnarray}
\cos\beta&=&\hat{n}_{L}\cdot\hat{n}_{\perp} \label{cbetadef}\\[2mm]
\cos\gamma&=&-\frac{\hat{n}_{L}\cdot\hat{q}_{3}}{
             |\hat{n}_{L}\times\hat{n}_{\perp}|}\nonumber
 \end{eqnarray}
where ($\hat a$ denotes a unit three-vector)
 \begin{itemize}
\item  $\hat n_L=-\hat n_Q$, with $\hat n_Q$  the direction of the
hadrons in the labframe,
\item
 $\hat{n}_\perp =\hat q_1\times \hat q_2$, the normal to the plane
defined by the momenta of particles 1 and 2.
 \end{itemize}
Note that the angle $\gamma$ defines a rotation around
$\hat{n}_{\perp}$ and determines
 the orientation of the three hadrons within
their production plane.
The definition of the angles $\beta$ and $\gamma$ is shown in fig.~1.

 Performing the integration over the
unobservable neutrino and the the unobservable Euler angle $\alpha$
we obtain the differential decay width for a polarized $\tau$
\cite{kumi1,kumi2}:
  \begin{eqnarray}
d\Gamma(\tau\rightarrow 3h)&=&
           \frac{G^{2}}{2\mt} \,
 \bigl(^{\cos^2\theta_{c}}_{\sin^2\theta_{c}}\bigr)
\, \left\{\sum_{X}\bar{L}_{X}W_{X}\right\}\times \label{diffrat}\\[3mm]
&&\frac{1}{(2\pi)^{5}}\frac{1}{64}
\frac{(\mt^{2}-Q^{2})^{2}}{\mt^{2}}\,
     \frac{dQ^{2}}{Q^{2}}\,ds_{1}\,ds_{2}
     \,\frac{d\gamma}{2\pi}\,
      \frac{d\cos\beta}{2}\,    \frac{d\cos\theta}{2}
      \nonumber
\end{eqnarray}
In (\ref{diffrat})
 we have defined the invariant masses in  the Dalitz plot
 $s_{i}=(q_{j}+q_{k})^{2}$
(where $i,j,k=1,2,3;\,i\neq j \neq k$)
 and the square
 of the invariant mass of the hadron system $Q^2\equiv
 (q_1+q_2+q_3)^2$.
The angle $\theta $ is related to the hadronic energy in the labframe
$E_h$ by \cite{kuwa,kumi1,kumi2}
 \begin{equation}
\cos\theta = \frac{\left(2x\mt^{2}-\mt^{2}-Q^{2}\right)}{
          (\mt^{2}-Q^{2}) \sqrt{1-4\mt^{2}/s}}
\label{cthdef}
  \end{equation}
with
  \begin{equation}
\hspace{0.1cm}  x  = 2\frac{E_{h}}{\sqrt{s}}
\hspace{1cm}  s = 4 E^{2}_{{beam}}
  \end{equation}
Another quantity depending on this energy $E_h$ is
  \begin{eqnarray}
\cos\psi &=&
       \frac{x(\mt^{2}+Q^{2})-2Q^{2}}{(\mt^{2}-Q^{2})
\sqrt{x^{2}-4Q^{2}/s}}
\label{psi}
  \end{eqnarray}
which will be of some interest in the subsequent discussion.
Finally
in the case where the spin zero part of the hadronic current $J^{\mu}$
is zero ($F_{4}=0$ in  (\ref{f1234})),
$ \sum_{X}\bar{L}_{X}W_{X}$
is given by a sum of nine terms
$\bar{L}_{X}W_{X}$
with $X\in\{A,B,C,D,E,F,G,H,I\}$
corresponding to nine density matrix elements of the hadonic
system in a spin one state.
One has \cite{kumi2}
\begin{equation}
  \begin{array}{lcrl}
\bar{L}_{A} &=& {2}/{3}&\ke\,+\,\kz
             \,+\,  1/3\,\keb\,(3\cthz-1)/2  \\[2mm]
\bar{L}_{B} &=& {2}/{3} & \ke\,+\,\kz
                 \,-\,  2/3\,\keb\,(3\cthz-1)/2  \\[2mm]
\bar{L}_{C} &=&  - 1/2&\keb\,\sthz\czchi\\[2mm]
\bar{L}_{D} &=&    1/2&   \keb\,\sthz\szchi\\[2mm]
\bar{L}_{E} &=&       &\kdb\,\cth\\[2mm]
\bar{L}_{F} &=&   1/2 &    \keb\,  \szth\cchi \label{ldef1}  \\[2mm]
\bar{L}_{G} &=&  -    &\kdb\,\sth\schi\\[2mm]
\bar{L}_{H} &=& -\,1/2&\keb\,\szth\schi   \\[2mm]
\bar{L}_{I} &=& -\,   &\kdb\,\sth\cchi
\end{array}\nonumber
\end{equation}
where
  \begin{eqnarray}
\ke &=&  1- P \cos\theta -
        ({\mt^{2}}/{Q^{2}})\, (1+  P \cos\theta)  \nonumber\\[1mm]
\kz &=&  ({\mt^{2}}/{Q^{2}})\, (1+  P \cos\theta)  \nonumber\\[1mm]
\kd &=&  1-P \cos\theta \nonumber \\[1mm]
\keb&=&  \ke \,(3\cpsiz-1)/2 - 3/2\,\kv\,\szpsi\nonumber \\[1mm]
\kzb&=& \kz\,\cpsi\,+\,\kv\,\spsi \label{kalldef}\\[1mm]
\kdb&=& \kd\,\cpsi\,-\,\kf\,\spsi \nonumber \\[1mm]
\kv &=&   \sqrt{{\mt^{2}}/{Q^{2}}}\,\, P \sin\theta \nonumber\\[1mm]
\kf &=&  \sqrt{\mt^{2}/Q^{2}}\,\,P\sin\theta \nonumber
  \end{eqnarray}
In (\ref{kalldef})  $P$  denotes the polarization of the $\tau$ in
the
laboratory frame
while $\theta$ and $\psi$ are defined in eqs. (\ref{cthdef},\ref{psi}).
For LEP-physics ($Z$-decay) $P$ is given by
$P=\frac{-2\,v_{\tau}a_{\tau}}{v_{\tau}^{2}+a_{\tau}^{2}}$ with
$v_{\tau}=-1+4\sin^{2}\theta_{W}$ and $a_{\tau}=-1$;
while for lower energies  $P$ vanishes.
In this case (for ARGUS, CLEO)  (\ref{kalldef})
 simplifies to
  \begin{eqnarray}
\ke &=&  1 - {\mt^{2}}/{Q^{2}}\,
=\,\frac{2\keb}{(3\cpsiz-1)}  \nonumber\\[1mm]
\kz &=&  {\mt^{2}}/{Q^{2}}\,=\,
\frac {\kzb}{\cpsi}  \label{kalldef2} \\[1mm]
\kd &=&  1\,=\, \frac{\kdb}{\cpsi} \nonumber
\end{eqnarray}
Note that the {\bf full} dependence on the $\tau$ polarization $P$,
 the hadron energy
(through $\theta$ and $\psi$)
and the angles $\beta$ and $\gamma$ is given  in eqs.~(\ref{ldef1})
to (\ref{kalldef2}).

The hadronic functions $W_{X}$
contain the dynamics of the hadronic decay and
depend in general on $s_{1},s_{2}$ and $Q^{2}$.
Let us recall that we
 are working in the hadronic rest frame with the $z
-$ and $x-$  axis aligned along $\hat{n}_{\perp}$ and
 $\hat{ q}_{3}$, respectively (see fig.~1).
The hadronic tensor $H^{\mu\nu}=J^{\mu}(J^{\nu })^{\dagger}$
(with $J^{\mu}$ given in  (\ref{f1234}))
 is calculated in this frame and the hadronic
structure functions $W_{X}$
 are linear combinations of density matrix elements
which are obtained from
  \begin{equation}
H^{\sigma\sigma^{\prime}}=\epsilon_{\mu}(\sigma)H^{\mu\nu}
                               \epsilon_{\nu}^{\ast}(\sigma^{\prime})
\label{hssdef}
  \end{equation}
where
  \begin{equation}
    \begin{array}{ll}
\epsilon_{\mu}(\pm)=\frac{1}{\sqrt{2}}\,(0;\pm1,-i,0)\hspace{1cm}&
\epsilon_{\mu}(0)  =(0;0,0,1)\\[2mm]
    \end{array}
\hspace{5mm}
\label{polvekdef}
  \end{equation}
 are the polarization vectors
for a   hadronic system in a spin one
 state defined with respect to the normal on the
three meson plane in the hadronic restframe.
The pure spin-one structure functions are
\begin{equation}
    \begin{array}{llrrlll}
W_{A} & =& &&\hspace{3mm} H^{++} + H^{--} &=&    H^{11}+H^{22} \\[2mm]
W_{B} & =& &&\hspace{3mm} H^{00}          &=&    H^{33} \\[2mm]
W_{C} & =& &-& ( \,H^{+-}+H^{-+})         &=& H^{11}-H^{22} \\[2mm]
W_{D} & =&          &i&( \,H^{+-}-H^{-+})&=& H^{12}+H^{21} \\[2mm]
W_{E} & =& &&\hspace{3mm}          H^{++}-H^{--}  &=&
-i\,(H^{12}-H^{21})\\[2mm]
W_{F} & =&  &-&  (H^{+0}+H^{0+}-H^{-0}-H^{0-})/ \sqrt{2}  &=&
H^{13}+H^{31}
 \\[2mm]
W_{G} & =&    &i&  (H^{+0}-H^{0+}-H^{-0}+H^{0-})/ \sqrt{2}  &=&-i\,
(H^{13}-H^{31})\\[2mm]
W_{H} & =&  &i&  (H^{+0}-H^{0+}+H^{-0}-H^{0-})/ \sqrt{2}  &=&
H^{23}+H^{32} \\[2mm]
W_{I} & =&  &&  (H^{+0}+H^{0+}+H^{-0}+H^{0-})/ \sqrt{2}   &=&
 -i\,(H^{23}-H^{32})\\[2mm]
    \end{array}
\label{hidef}
\end{equation}
The  r.h.s of eqs. (\ref{hidef}) refers to the cartesian
components of
$H^{\mu\nu}$.
The structure
functions can thus be expressed in terms of the formfactors
$F_{i}$
as follows \cite{kumi2}
  \begin{eqnarray}  \hspace{3mm}
W_{A}  &=&   \hspace{3mm}(x_{1}^{2}+x_{3}^{2})\,|F_{1}|^{2}
           +(x_{2}^{2}+x_{3}^{2})\,|F_{2}|^{2}
 +2(x_{1}x_{2}-x_{3}^{2})\,\mbox{Re}\left(F_{1}F^{\ast}_{2}\right)
                                   \nonumber \\[5mm]
W_{B}  &=& \hspace{3mm} x_{4}^{2}|F_{3}|^{2}
                                   \nonumber \\[5mm]
W_{C}  &=&  \hspace{3mm} (x_{1}^{2}-x_{3}^{2})\,|F_{1}|^{2}
           +(x_{2}^{2}-x_{3}^{2})\,|F_{2}|^{2}
           +2(x_{1}x_{2}+x_{3}^{2})
           \,\mbox{Re}\left(F_{1}F^{\ast}_{2}\right)
                                   \nonumber \\[5mm]
W_{D}  &=&  \hspace{3mm}2\left[ x_{1}x_{3}\,|F_{1}|^{2}
           -x_{2}x_{3}\,|F_{2}|^{2}
           +x_{3}(x_{2}-x_{1})\,
           \mbox{Re}\left(F_{1}F^{\ast}_{2}\right)\right]
                                   \nonumber \\[5mm]
W_{E}  &=& -2x_{3}(x_{1}+x_{2})\,\mbox{Im}\left(F_{1}
                    F^{\ast}_{2} \right)\nonumber \\[5mm]
W_{F}  &=&  \hspace{3mm}
          2x_{4}\left[x_{1}\,\mbox{Im}\left(F_{1}F^{\ast}_{3}\right)
          + x_{2}\,\mbox{Im}\left(F_{2}F^{\ast}_{3}\right)\right]
                                   \nonumber \\[2mm]
W_{G}  &=&- 2x_{4}\left[x_{1}\,\mbox{Re}\left(F_{1}F^{\ast}_{3}\right)
              + x_{2}\,\mbox{Re}\left(F_{2}F^{\ast}_{3}\right)\right]
                                   \nonumber \\[2mm]
W_{H}  &=& \hspace{3mm}
      2x_{3}x_{4}\left[\,\mbox{Im}\left(F_{1}F^{\ast}_{3}\right)
                     -\,\mbox{Im}\left(F_{2}F^{\ast}_{3}\right)\right]
                                   \label{www} \\[2mm]
W_{I}  &=&- 2x_{3}x_{4}\left[\,\mbox{Re}\left(F_{1}F^{\ast}_{3}\right)
                     -\,\mbox{Re}\left(F_{2}F^{\ast}_{3}\right)\right]
                                   \nonumber \\[2mm]
\end{eqnarray}
where
$x_{i}$ are functions of the Dalitz plot variables and $Q^{2}$.
One has
\begin{eqnarray}
x_{1}&=q_{1}^{x}-q_{3}^{x}\nonumber\\[1mm]
x_{2}&=q_{2}^{x}-q_{3}^{x}\nonumber\\[1mm]
x_{3}&=q_{1}^{y}=-q_{2}^{y}\\[1mm]
x_{4}&=\sqrt{Q^{2}}x_{3}q_{3}^{x}\nonumber
   \end{eqnarray}
Here $E_{i}$ and $q_{i}$ refers to the components of the hadron
momenta  in the hadronic
rest frame with
  \begin{eqnarray}
E_{i}&=&\frac{Q^{2}-s_{i}+m_{i}^{2}}{2\sqrt{Q^{2}}}\nonumber\\[1mm]
q_{3}^{x}&=&\sqrt{E_{3}^{2}-m_{3}^{2}}\nonumber\\[1mm]
q_{2}^{x}&=& (2E_{2}E_{3}-s_{1}
+m_{2}^{2}+m_{3}^{2})/(2q_{3}^{x})\nonumber\\[1mm]
q_{1}^{x}&=&
(2E_{1}E_{3}-s_{2}+m_{1}^{2}+m_{3}^{2})/(2q_{3}^{x})\nonumber\\[1mm]
q_{2}^{y}&=&-\sqrt{E_{2}^{2}-(q_{2}^{x})^{2}-m_{2}^{2}}\nonumber\\[1mm]
q_{1}^{y}&=&\,\,\,\sqrt{E_{1}^{2}-(q_{1}^{x})^{2}-m_{1}^{2}}
            \,=-q_{2}^{y}\nonumber
  \end{eqnarray}
Note that the structure
 functions $W_{B,F,G,H,I}$ are related to the anomaly
formfactor $F_{3}$.
In the following we will consider these structure functions
in more detail.
\section{Def\/inition of Moments}\label{diss}
Equation (\ref{diffrat})
provides the full description of the angular distribution
of the decay products from a single polarized $\tau$.
They reveal that the measurement of
the structure functions
 $W_{i}$ and therefore the measurement of the anomaly
formfactor $F_{3}$ is possible in currently ongoing high statistics
experiments.
In the following we will concentrate on the $s_{1},s_{2}$ integrated
stucture functions
\begin{equation}
w_i(Q^2)\equiv \int ds_1ds_2 W_i(Q^2,s_1,s_2)
\label{kwdef}
\end{equation}
A possible strategy to isolate the various structure functions in
(\ref{diffrat}) is to take suitable moments on the differential decay
distribution \cite{kumi2}.
Let us define\footnote{
Note that these moments differ from the moments defined
in
\cite{kumi2} by the factor $R_{^{c}_{s}}$ defined in  (\ref{rdef}).}
  \begin{equation}
\langle  f(\beta\,,\gamma ) \rangle =\int
\frac{8\pi\sqrt {Q^2} d\Gamma(\tau\to\nu_\tau\, 3h)}
{dQ^2\,d\cos\theta\,d\gamma\, d\cos\beta}\,
\,f(\beta\,,\gamma )\, \frac{d\cos\beta}{2}\,\frac{d\gamma}{2\pi}
\label{moments}
  \end{equation}
which yields
\begin{eqnarray}
\langle 1 \rangle &=& \hspace{3mm}
R_{^{c}_{s}}(Q^2)\,\,\,(2 \ke+3\kz )\,(w_{A}+w_{B})\nonumber\\[2mm]
\langle (3\cos^{2}\beta-1)/2 \rangle &=&
\hspace{3mm}{R_{^{c}_{s}}(Q^2)}\,\,\,\frac{1}{5}\,
\keb\,(w_{A}-2w_{B})\nonumber\\[2mm]
\langle \cos 2\gamma \rangle &=&
-{R_{^{c}_{s}}(Q^2)}\,\,\,\frac{1}{2}\, \keb\,w_{C}\nonumber\\[2mm]
\langle \sin 2\gamma \rangle &=&
\hspace{3mm}{R_{^{c}_{s}}(Q^2)}
\,\,\,\frac{1}{2}\, \keb\,w_{D}\nonumber\\[2mm]
\langle \cos\beta \rangle &=&
\hspace{3mm}\,{R_{^{c}_{s}}(Q^2)}\,\,\, \kdb\,w_{E} \nonumber\\[2mm]
\langle \sin 2\beta \cos\gamma \rangle &=&
\hspace{3mm}R_{^{c}_{s}}
\,\,\,\frac{2}{5}\, \keb\,w_{F}\label{mof3}\\[2mm]
\langle \sin\beta \sin\gamma \rangle &=&
-\,R_{^{c}_{s}}(Q^2)\,\,\,\kdb\,w_{G}\nonumber\\[2mm]
\langle \sin 2\beta \sin\gamma \rangle &=&
-R_{^{c}_{s}}(Q^2)\,\,\,\frac{2}{5} \keb\,w_{H} \nonumber\\[2mm]
\langle \sin\beta \cos\gamma \rangle &=&
-\,R_{^{c}_{s}}(Q^2)\kdb\,w_{I}\nonumber
\end{eqnarray}
where the function $R_{^{c}_{s}}(Q^2)$ has been defined by
\begin{equation}
R_{^{c}_{s}}(Q^2)=
\frac{G^{2}}{12\mt{^3}}
\,\bigl(^{\cos^2\theta_{c}}_{\sin^2\theta_{c}}\bigr)
 \frac{1}{(4\pi)^{5}}
\frac{(\mt^{2}-Q^{2})^{2}}{\sqrt{Q^2}}
\label{rdef}
\end{equation}
Some comment are in order here:
\begin{itemize}
\item
First note that after integration over the angles
$\beta$ and $\gamma$ the preceding expressions
are
still dependent on $P$ and $E_h$ (through $\theta$ and $\psi$) while
the hadronic structure functions $w_{X}$ are functions of $Q^{2}$.
\item The sum $w_{A}+w_{B}$ is
 closely related to the spin one
part of the spectral function:
  \begin{equation}
\rho_{1}(Q^{2})=\frac{1}{6}\frac{1}{(4\pi)^{4}}\frac{1}{Q^{4}}\,
                (w_{A}+w_{B})\label{st1}
\end{equation}
and we obtain the standard form for the total width
  \begin{eqnarray}
\hspace{-1cm}
\Gamma(\tau\rightarrow \nu_\tau\,3h)&=&
           \frac{G^{2}}{8\pi\mt}
           \,\bigl(^{\cos^2\theta_{c}}_{\sin^2\theta_{c}}\bigr)
\int dQ^{2} \,(\mt^{2}-Q^{2})^{2}
\left(1+\frac{2Q^{2}}{\mt^{2}}\right)\rho_{1}(Q^2) \\[2mm]
&=&
\int \frac{dQ^2}{\sqrt{Q^2}}
R_{^{c}_{s}}(Q^2) \big( \frac{\mt^2+2Q^2}{Q^2}\big)(w_A+w_B)
\nonumber
\end{eqnarray}
\end{itemize}
In our figures in section (\ref{numer}) we will present the functions
$R_{^{c}_{s}}(Q^2)w_X(Q^2)$
as well as numerical results for the hadronic
structure functions itself.

In the next section we present an explicit parametrization of the
formfactors which are used in our numerical simulations in order to
test whether the anomaly can be measured experimentally.

\section{Formfactors} \label{form}
In a recent paper \cite{demi} we have given an explicit parametrization
of the form factors and compared successfully with measured widths.
The physical idea behind the model for the formfactors can be resumed
to:
\begin{itemize}
\item In the chiral limit the formfactors are normalized to the
$U(3)_L\times U(3)_R$ chiral model.
\item Meson-vertices are independent of momentum.
\item The full momentum dependence  is given by
Breit-Wigner propagators  of the resonances occuring in
the different channels.
Resonances occur either in $Q^{2}$ which are then three
body resonances or in $s_{i}$ which are two body resonances.
\end{itemize}
Now we present our parametrization for the
 formfactors $F_{i}$ defined in (\ref{f1234}) which fulfill all these
requirements.

 First we
present the formfactors induced by the anomaly
 \begin{eqnarray}
F_3^{(\eta\pi^-\pi^0)}(s_1,Q^2) =&{1\over 2\sqrt 6\pi^2f^3_\pi}
T^{(2)}_\rho[Q^2]T^{(1)}_\rho[s_1] \\[2mm]
F_3^{(K^-\pi^-K^+)}(s_1,s_2,Q^2) =&{-1\over 2\sqrt 2\pi^2f^3_\pi}
 T^{(2)}_{\rho}[Q^2]\,T_{\rho K^*} (s_2,s_1,\alpha)\\[2mm]
F_3^{(K^-\pi^-\pi^+)}(s_1,s_2,Q^2) =&{1\over 2\sqrt 2\pi^2f^3_\pi}
T^{(1,2)}_{K^*}[Q^2]\,T_{\rho K^*}(s_1,s_2,\alpha)
\label{f3kpipi}
\end{eqnarray}
The parametrization of the $\eta\pi^-\pi^0$ channel is obtained from
$e^++e^-$-data via CVC \cite{pich3,demi}.
It is given as a product of two functions describing the
resonances in $Q^2$ and $s_i$.  The same $Q^2$ dependence
($T^{(2)}_\rho[Q^2]$) can be used
 for the $K^-\pi^-K^+$-channel. Of course the two-body
channels have to be modified since they involve  a $\rho$ and a $K^*$
as well,
we have included these
contributions in $T_{\rho K^*} (s_2,s_1,\alpha)$\cite{demi}.
The same function $T_{\rho K^*} $
also enters in the $K^-\pi^-\pi^+$ channel.
Unfortunately
nothing is known experimentally
on the three-body resonances in this channel.
In \cite{demi}
only the $K^*$ resonance ($T^{(1)}_{K^*}$) has been included. In our
numerical results we will also use a different parametrization
including more $\Delta S=1$ vector resonances.

Second the axialvector current induces two formfactors $F_1$ and $F_2$
for the $K^{-}\pi^{-}K^{+}$ and $K^{-}\pi^{-}\pi^{+}$ channels with the
following parametrization
\begin{eqnarray}
 F^{(K^-\pi^-K^+)}_{1}(s_2,Q^2)&={-\sqrt 2 \over 3f_\pi}BW_{A_1}[Q^2]\,
T^{(1)}_\rho[s_2]\\[2mm]
 F^{(K^-\pi^-K^+)}_{2}(s_1,Q^2)&={-\sqrt 2 \over 3f_\pi}BW_{A_1}[Q^2]\,
T^{(1)}_{K^*}[s_1]
\end{eqnarray}
and
\begin{eqnarray}
 F^{(K^-\pi^-\pi^+)}_{1}(s_2,Q^2)&={-\sqrt 2 \over
 3f_\pi}\widehat{BW}_{K_1}[Q^2]\,
T^{(1)}_{K^*}[s_2]\\[2mm]
 F^{(K^-\pi^-\pi^+)}_{2}(s_1,Q^2)&={-\sqrt 2 \over
 3f_\pi}\widehat{BW}_{K_1}[Q^2]\,
T^{(1)}_\rho[s_1]
 \end{eqnarray}
Note that $G$-parity forbids the axialvector current to contribute to
the
$\eta\pi^-\pi^0$-channel.
In the axialvector  channel we assume the dominance
of a resonance in each channel, i.e. the $A_1$ and the $K_1$
in the $\Delta S=0$ and $\Delta S=1$ channel, respectively.
The two-body
channels are  again parametrized by the functions $T^{(1)}_\rho$
and
$T^{(1)}_{K^*}$.
We have moved explicit expressions of these functions and
all numerical parameters (taken from \cite{demi}) to appendix A.\\

\section{Numerical results }\label{numer}
In this section we
 will present
 numerical results for $R_{^{c}_{s}}(Q^{2)} \cdot w_{X}(Q^{2})$
(defined in (\ref{kwdef},\ref{rdef}))
as well as for
the hadronic structure functions $w_{X}(Q^{2})$ separately for
the different decay channels.
We prefer to
present both $R_{^{c}_{s}}\cdot w_{X}$ and $w_{X}$ in order to
show the effect of the phase space
(included in the function $R_{^{c}_{s}}$)
 while
the hadronic physics is more visible in the plots for $w_{X}$ alone.
 Although
by integrating over $s_1$ and $s_2$ we have lost information on the
resonances in the two body decays  we observe still
interesting structures.

Let us start with the Cabibbo
allowed decay $\tau\to\nu_\tau+\eta\pi^-\pi^0$.
As mentioned before this channel has a vanishing contribution from the
axialvector current ($G$-parity)
which implies that only $w_{B}$ is different from zero.
A comparison of the data and our prediction for $R_{c}\cdot w_{B}$ and
$w_{B}$ in fig.~(2a,b)  would
be highly interesting, especially a confirmation of higher lying $\rho$
resonances in $T_{\rho}^{(2)}(Q^{2})$ (the shoulder at $Q^2=3$GeV$^2$
in fig.~(2a,b)).

Next process is the Cabibbo allowed decay
$\tau\to\nu_\tau K^-\pi^{-}K^+$
which has contributions
from both the axialvector and vector currents.
Therefore all nine  structure functions  are different from
zero.
In fig. 3a we present the
structure function combinations obtained from the
 $\langle 1 \rangle$ and $\langle
(3\cos^{2}\beta-1)/2)$  moment.
 Note that a measurement of the differential decay width
(proportional to $ \langle 1\rangle$) is not
enough to separate $w_A$ and $w_B$.
We observe a sizeable
effect of $w_B$ which makes a determination of $F_3$
possible.
 Note this sizeable effect is due to heavy $\rho^{\prime}$ resonances
in the anomaly channel, which existence is predicted
from the description of $e^{+}e^{-}\rightarrow\eta\pi\pi$.
In order to get
a feeling of the effects of the phase space in this channel
 we present the combinations
$w_A+w_B$
and $w_A-2w_B$ as well as the structure functions $w_{A}$ and
$w_{B}$ in fig.~3b.
The moments which measure the interference of
the axialvector  and vector-currents are presented
in fig.~3c.
 The size of these moments is comparable to those in fig.~3a
and the  very
peculiar shape would make them  measurable too.
For completeness we present the remaining moments in fig.~3d.

Finally we discuss  the Cabibbo suppressed decay
$\tau\rightarrow
\nu_{\tau}K^{-}\pi^{-}\pi^{+}$ in figs. 4a-d for the parametrization
with
$T_{K^{*}}^{(1)}$ in  (\ref{f3kpipi}).
Note that although
this decay is Cabibbo-suppressed the moments are
comparable
 in size to the $K^-\pi^{+}K^+$ case (suppression due to the
mass of the supplementary kaon in the phase space is comparable to
the Cabibbo-suppression).
 All moments
of figs. 4a-d  have a shape which shows the strong presence of
the $K_1$
resonance
 in the axial channel. A measurement of the structure functions
related to the anomaly seems very hard since
 $F_{3}$  is very small in this parametrization.
Of course this unfavorable result could have been deduced since the
contribution of the anomaly to the rate, as computed in \cite{demi},
was of the order of $1\%$. However, we should note that  our
parametrization
of the anomaly form factor \cite{demi} includes only a $K^*$,
which can never
 be on mass shell,
and therefore produces no strong enhancement. On the other
hand in this channel we have no CVC prediction which could tell us
if heavier resonances are present in this channel.
In order to get a feeling
for possible effects of heavier $K^{*}$ resonaces
 we propose   the following parametrization which is $\rho$-channel
inspired (see eq.\ref{trho}).
\begin{equation}
 T^{(2)}_{K^*}[s]={1\over 1+\beta+\delta}\Biggl\{
BW_{K*(1680)}[s]+\beta BW_{K^*(1410)}[s]+\delta  BW_{K^* }[s]\Biggr\}
\label{k2}
 \end{equation}
$$
 \begin{array}{lll}
 \delta=-26 &  m_{K^* }=0.892\ {\rm GeV}\ & \Gamma_{K^* }=0.050\
 {\rm   GeV } \\[2mm]
 \beta =6.5 &  m_{K^*(1410)}=1.412\
 {\rm GeV}  & \Gamma_{K^*(1410)}=0.227\
 {\rm  GeV}   \\[2mm]
            &   m_{K^*(1680)}=1.714\
            {\rm GeV}&\ \Gamma_{K^*(1680)}=0.323\
 {\rm  GeV}    \\[2mm]
\end{array}
$$
With this parametrization we obtain the results in figs. 5a-c.
Note that fig. 4d is not changed.

For completeness we present the results of the total decay width
$\Gamma^{\eta\pi^{-}\pi^{0}}$, $\Gamma^{K^{-}\pi^{-}K^{+}}$ and
$\Gamma^{K^{-}\pi^{-}\pi^{+}}$ normalized to the electronic width
$\Gamma_{e}$ ($\Gamma_{e}/ \Gamma_{tot}\approx 18\%$). One
has\cite{demi}\\[2mm]
$$ \matrix{ {\rm Channel}\ (abc)&&{\Gamma^{(abc)}\over \Gamma_e}&&
 {\mbox{Contribution from $F_{3}$ in \%}}&\cr
&&&&&\cr
\eta\pi^-\pi^0 &&0.0108&&100\,\,\%&\cr
K^-\pi^-K^+&&0.0061&&39.3\,\,\%&\cr
K^-\pi^-\pi^+&&0.0316&&1.1\,\,\%
\hspace{1cm}\mbox{with $T_{K^{*}}^{(1)}$ }&\cr
K^-\pi^-\pi^+&&0.0325&&4.2\,\,\%
\hspace{1cm}\mbox{with $T_{K^{*}}^{(2)}$}&\cr
&&&&&\cr}$$

In view of this result we urge our experimental colleagues to study
carefully this Cabibbo suppressed channel.
\section{ Conclusions}\label{conc}
In this paper we have proposed to measure moments (eq. (\ref{moments}))
which allow to determine
quantitatively the
contribution of the Wess-Zumino-anomaly to $\tau$ decays into
three mesons. We have considered the channels $\eta\pi^-\pi^0$,
$K^-\pi^-K^{+}$ and $K^-\pi^-\pi^+$.
We have shown that  measuring the unique moment of
$\eta\pi^-\pi^0$ channel allows to verify the CVC prediction,
especially the presence of heavy $\rho$ excitations observed in
$e^++e^-$ data \cite{DMS}.
In the
$K^-\pi^{-}K^+$ channel we can define much more moments because
of the interference
of the anomaly with the axial vector contributions.
In our prediction Fig.~(2)
the effect of the heavier $\rho$ is again
clearly seen.

The interest of the analysis of the $K^-\pi^-\pi^+$ is twofold:
we learn
something about resonances, first in the axialvector channel
and second in the vector channel.
We noted that
in contradistinction to the Cabibbo allowed decays the vector
channel
for Cabibbo
 suppressed cannot be predicted through CVC from $e^++e^-$-data.\\

\appendix
\section{Parameters used in the formfactors}\label{appa}
As stated in section \ref{form} the formfactors are dominated by
resonances. The effects of these resonances are described by
functions  $BW_P(Q^2)$, which are
normalized ($BW(0)=1$) Breit-Wigner propagators.
We will use two kind of Breit-Wigners:
\begin{itemize}
\item energy-dependent width
\begin{equation}
 BW_{R}[s]\equiv {-M^2_R\over [s-M^2_R+i\sqrt s \Gamma_R(s)]}
\end{equation}
The energy-dependent width ($\Gamma_R(s)$) is computed from its
usual definition.
\begin{equation}
\Gamma_R(s)={1\over 2\sqrt s}|M_{R\to  f_i}|^2d\Phi\delta
(Q-\sum p_i)
\qquad \Gamma_R(s)_{s=M_R^2}=\Gamma_R
\end{equation}
\item constant width
\begin{equation}
 \widehat{BW}_{R}[s]\equiv
 {-M^2_R+iM_R\Gamma_R\over [s-M^2_R+iM_R \Gamma_R]}
\end{equation}
\end{itemize}
First we define the parameters which arise in the
axialvector three-body channel:
In the Breit-Wigner of the $A_1$ we use
$$ m_{A_1}= 1.251\ {\rm GeV} \quad \Gamma_{A_1}= 0.599\ {\rm GeV}$$
\mbox{}
\begin{equation}
\sqrt s\Gamma_{A_1}(s)=m_{A_1}\Gamma_{A_1} {g(s)\over g(m^2_{A_1})}
\end{equation}
where the function $g(s)$ has been  calculated in \cite{kuhnsanta}
\begin{equation}
g(s)=\pmatrix{4.1(s-9m^2_\pi)^3(1-3.3(s-9m^2_\pi)+5.8(s-9m^2_\pi)^2)
& {\rm if}\ s<(m_\rho+m_\pi)^2\cr
 &&\cr
 s(1.623 +{10.38\over s}-{9.32\over s^2}
 +{0.65\over s^3})& {\rm else}\cr}
\end{equation}
In this equation all masses and $\sqrt s$ are expressed in GeV. \\
 In case of the $K_1$ resonance we use a constant width Breit-Wigner
$\widehat B_{K_1}(s)$
with
$$ m_{K_1}= 1.402 \ {\rm GeV}\quad \Gamma_{K_1}= 0.174 \,{\rm GeV}$$
This is because the decay of the $K_1$ is  experimentally not  well
known.

The Cabibbo allowed vector formfactor is obtained from CVC and yields
\cite{DMS}
 \begin{equation}
T^{(2)}_{\rho}[s]={1\over 1+\beta+\delta}\Biggl\{
BW_{\rho ''}[s]+\beta BW_{\rho '}[s]+\delta  BW_{\rho }[s]\Biggr\}
\label{trho}
 \end{equation}
$$
 \begin{array}{lll}
 \delta=-26 &  m_{\rho }=0.773\ {\rm GeV}\ & \Gamma_{\rho }=0.145\
 {\rm   GeV } \\[2mm]
 \beta =6.5 &  m_{\rho '}=1.500\ {\rm GeV}  & \Gamma_{\rho '}=0.220\
 {\rm  GeV}   \\[2mm]
          &   m_{\rho ''}=1.750\ {\rm GeV}&\ \Gamma_{\rho ''}=0.120\
 {\rm  GeV}    \\[2mm]
\end{array}
$$
For its $\Delta S=1$ we propose (no experimental data)
either
 \begin{equation}
 T^{(1)}_{K^*}[s]\equiv BW_{K^{*}}[s]\label{k3}
 \end{equation}
$$
m_{K^*}= 0.892\ {\rm GeV}\quad \Gamma_{K^*}=0.051\ {\rm GeV}
$$
or the function $ T^{(2)}_{K^*}[s]$ defined in (\ref{k2}).

Finally we define
the functions describing the resonances in the two-body
channel \cite{kuhnsanta}
 \begin{equation}
T^{(1)}_{\rho}[s]={1\over 1+\beta}\Biggl\{
BW_{\rho}[s]+\beta BW_{\rho '}[s]\Biggr\}
\label{t1r} \end{equation}
$$
 \begin{array}{lll}
   & m_{\rho }=0.773\
   {\rm GeV}\ & \Gamma_{\rho }=0.145\ {\rm GeV} \\[2mm]
\beta=-0.145\
  & m_{\rho '}=1.370\ {\rm GeV}
  & \Gamma_{\rho '}=0.510\ {\rm GeV}\\[2mm]
 \end{array}
$$
The function $
 T^{(1)}_{K^*}[s]$ has been defined in (\ref{k3}). Last but not
least we define
the function which enters the anomaly two-body channel
 \begin{equation}
 T_{\rho K^*} (s_1,s_2,\alpha )={T^{(1)}_\rho [s_1]+\alpha T_{K^*}[s_2]
\over 1+\alpha}
 \end{equation}
where $T^{(1)}_{\rho}$
is given in (\ref{t1r}) and $\alpha=-0.2$ \cite{demi}.

\newpage
\def\app#1#2#3{{\it Act. Phys. Pol. }{\bf B #1} (#2) #3}
\def\apa#1#2#3{{\it Act. Phys. Austr.}{\bf#1} (#2) #3}
\def\lhc{Proc. LHC Workshop, CERN 90-10}
\def\npb#1#2#3{{\it Nucl. Phys. }{\bf B #1} (#2) #3}
\def\plb#1#2#3{{\it Phys. Lett. }{\bf B #1} (#2) #3}
\def\prd#1#2#3{{\it Phys. Rev. }{\bf D #1} (#2) #3}
\def\prl#1#2#3{{\it Phys. Rev. Lett. }{\bf #1} (#2) #3}
\def\prc#1#2#3{{\it Phys. Reports }{\bf C #1} (#2) #3}
\def\cpc#1#2#3{{\it Comp. Phys. Commun. }{\bf #1} (#2) #3}
\def\nim#1#2#3{{\it Nucl. Inst. Meth. }{\bf #1} (#2) #3}
\def\pr#1#2#3{{\it Phys. Reports }{\bf #1} (#2) #3}
\def\sovnp#1#2#3{{\it Sov. J. Nucl. Phys. }{\bf #1} (#2) #3}
\def\jl#1#2#3{{\it JETP Lett. }{\bf #1} (#2) #3}
\def\jet#1#2#3{{\it JETP Lett. }{\bf #1} (#2) #3}
\def\zpc#1#2#3{{\it Z. Phys. }{\bf C #1} (#2) #3}
\def\ptp#1#2#3{{\it Prog.~Theor.~Phys.~}{\bf #1} (#2) #3}
\def\nca#1#2#3{{\it Nouvo~Cim.~}{\bf #1A} (#2) #3}
\sloppy
\raggedright

\newpage
\noindent
{\bf \hspace{-5mm} Figure captions}\\[2mm]
\begin{itemize}
\item[{\bf Fig. 1}]
Definition of the polar angle $\beta$ and the azimuthal angle
$\gamma$.
$\beta$ denotes the angle between $\vec{n}_{\perp}$
and $\vec{n}_{L}$.
$\gamma$ denotes the angle between the
 $(\vec{n}_{L},\vec{n}_{\perp})$ plane
 and the
$ (\vec{n}_{\perp},\hat{q}_{3})$-plane.
\item[{\bf Fig. 2}]
a) $Q^{2}$ dependence of $w_{B}\cdot R_{C}$
 for the decay channel $\eta\pi^{-}\pi^{0}$\\
b) $Q^{2}$ dependence of $w_{B}$
 for the decay channel $\eta\pi^{-}\pi^{0}$
\item[{\bf Fig. 3}]
a) $Q^{2}$ dependence of
    $(w_{A}+w_{B})\cdot R_{C}$,
    $(w_{A}-2 w_{B})\cdot R_{C}$,
    $ w_{A}\cdot R_{C}$ and $w_{B}\cdot R_{C}$
 for the decay channel $K^{-}\pi^{-}K^{+}$\\
b) $Q^{2}$ dependence of
    $(w_{A}+w_{B})$,
    $(w_{A}-2 w_{B})$,
    $ w_{A}$ and $w_{B}$
 for the decay channel $K^{-}\pi^{-}K^{+}$\\
c) $Q^{2}$ dependence of
    $w_{F}\cdot R_{C}$,
    $w_{G}\cdot R_{C}$,
    $w_{H}\cdot R_{C}$
and $w_{I}\cdot R_{C}$
 for the decay channel $K^{-}\pi^{-}K^{+}$\\
d) $Q^{2}$ dependence of
    $w_{C}\cdot R_{C}$,
    $w_{D}\cdot R_{C}$
and $w_{E}\cdot R_{C}$
 for the decay channel $K^{-}\pi^{-}K^{+}$
\item[{\bf Fig. 4}]
a) $Q^{2}$ dependence of
    $(w_{A}+w_{B})\cdot R_{S}$,
    $(w_{A}-2 w_{B})\cdot R_{S}$
     and $w_{B}\cdot R_{S}$
 for the decay channel $K^{-}\pi^{-}\pi^{+}$
 with the parametrization
$T_{K^{*}}^{(1)}$ in (\ref{k3})\\
b) $Q^{2}$ dependence of
    $(w_{A}+w_{B})$,
    $(w_{A}-2 w_{B})$
     and $w_{B}$
 for the decay channel $K^{-}\pi^{-}\pi^{+}$
 with the parametrization
$T_{K^{*}}^{(1)}$ in (\ref{k3})\\
c) $Q^{2}$ dependence of
    $w_{F}\cdot R_{S}$,
    $w_{G}\cdot R_{S}$,
    $w_{H}\cdot R_{S}$
 and $w_{I}\cdot R_{S}$
 for the decay channel $K^{-}\pi^{-}\pi^{+}$
  with the parametrization
$T_{K^{*}}^{(1)}$ in (\ref{k3})\\
d) $Q^{2}$ dependence of
    $w_{S}\cdot R_{S}$,
    $w_{D}\cdot R_{S}$
 and $w_{E}\cdot R_{S}$
 for the  decay channel $K^{-}\pi^{-}\pi^{+}$
  with the parametrization
$T_{K^{*}}^{(1)}$ in (\ref{k3})
 \item[{\bf Fig. 5}]
a-c) same as Fig. 4 with
the parametrization $T_{K^{*}}^{(2)}$ in (\ref{k2}).
\end{itemize}


\begin{thebibliography}{99}
\bibitem{WZ} J.~Wess and B.~Zumino, {\bf Phys.~Lett.~ B37B} 95 (1971)
\bibitem{kramer} G.~Kramer,
        W.~F.~Palmer and S.~Pinsky, {\bf  Phys.~Rev.~D30} 89
        (1984)  \\
 G.~Kramer, W.~F.~Palmer, {\bf  Z.~Phys.~C25} 195 (1984) \\
 G.~Kramer, W.~F.~Palmer, {\bf  Z.~Phys.~C39} 423 (1988)
\bibitem{pich3} A.~Pich, {\bf Phys.~Lett.~B196} 561 (1987)
\bibitem{demi} R.~Decker,
        E.~Mirkes, R.~Sauer and Z.~Was, Karlsruhe preprint TTP92--25,
to be published in {\bf Z. Phys. C}, in press.
\bibitem{CLEO} CLEO preprint,
        {\bf Measurement of $\tau$ Decays Involving $\eta$
        Mesons,} CLEO 92-7.
\bibitem{kumi1}
 J.H. K\"uhn and E. Mirkes,
  \plb{286}{1992}{381}
\bibitem{kumi2} J.H. K\"uhn and E. Mirkes, Karlsruhe preprint TTP92-20
              to appear in {\bf Z. Phys. C}, in press.
\bibitem{kuwa} J.H.\ K\"uhn
        and F.\ Wagner, Nucl.\ Phys.\ B236 (1984) 16.
\bibitem{kuhnsanta} J.~H.~K\"uhn and
        A.~Santamaria, {\bf Z.~Phys.~C48} 445 (1990)
\bibitem{DMS}
DM2 Collaboration,
A.~Antonelli et al., {\bf Phys.~Lett.~B212} 133 (1988)\\
J.~J.~ Gomez-Cadenas,
M.~C. Gonzales-Garcia and A.~Pich, {\bf Phys.~Rev.~D42 }
3093 (1990)
\end{thebibliography}
\end{document}